\documentclass[aps,pra,amsmath,superscriptaddress,reprint,floatfix,notitlepage,balancelastpage,twocolumn,showpacs,reprint]{revtex4}
\usepackage[colorlinks,urlcolor=red,citecolor=red]{hyperref}
\usepackage{graphicx} 
\usepackage{epstopdf}
\usepackage{amsthm} 

\usepackage{tikz}

\newcommand{\gv}[1]{\ensuremath{\mbox{\boldmath$ #1 $}}}
\newcommand{\abs}[1]{\left| #1 \right|} 
\newcommand{\f}[2]{\frac{#1}{#2}} 
\renewcommand{\d}[2]{\frac{d #1}{d #2}} 

\newcommand{\grad}[1]{\gv{\nabla} #1} 
\let\baraccent=\= 
\renewcommand{\=}[1]{\stackrel{#1}{=}} 

\theoremstyle{definition}

\theoremstyle{remark}

\newcommand{\uparrowp}{{\uparrow'}}
\newcommand{\downarrowp}{{\downarrow'}}

\newcommand{\vtrap}{V_{\rm trap}}
\newcommand{\abohr}{a_{\rm B}}

\newcommand{\curH}{{\cal H}}
\newcommand{\curN}{{\cal N}}

\newcommand{\phdag}{{\phantom{\dagger}}}

\newcommand{\curO}{{\cal O}}

\newcommand{\Omegah}{\hat{\Omega}}

\newcommand{\bk}{{\bf k}}

\newcommand{\bp}{{\bf p}}

\newcommand{\zh}{\hat{z}}
\newcommand{\xh}{\hat{x}}
\newcommand{\yh}{\hat{y}}

\newcommand{\Hh}{\hat{H}}
\newcommand{\Uh}{\hat{U}}

\newcommand{\deltah}{\hat{\delta}}
\newcommand{\psih}{\hat{\psi}}
\newcommand{\Psih}{\hat{\Psi}}

\newcommand{\br}{{\bf r}}

\newcommand{\be}{\begin{equation}}
\newcommand{\ee}{\end{equation}}
\newcommand{\bea}{\begin{eqnarray}}
\newcommand{\eea}{\end{eqnarray}}
\newcommand{\bse}{\begin{subequations}}
\newcommand{\ese}{\end{subequations}}

\begin{document}

\title{Density profiles and collective modes of a Bose-Einstein condensate with light-induced
spin-orbit coupling}
\author{Qin-Qin L\"u}
\author{Daniel E. Sheehy}
\email{sheehy@lsu.edu}
\affiliation{Department of Physics and Astronomy, Louisiana State University, Baton Rouge, LA, 70803, USA}
\date{June 24, 2013}
\begin{abstract} The phases of a Bose-Einstein condensate (BEC) with light-induced
spin-orbit coupling (SOC) are studied within the mean-field approximation.  
The mixed BEC phase, in which the system condenses in a superposition of two plane wave states,
is found to be stable for sufficiently small light-atom coupling, becoming unstable in
a continuous fashion with increasing light-atom coupling.  The structure of the phase diagram
at fixed chemical potential for bosons with SOC is shown to imply an unusual density dependence for a trapped
mixed BEC phase, with the density of one dressed spin state {\em increasing\/} with 
increasing radius, providing a unique experimental signature of this state.  The 
collective Bogoliubov sound mode is shown to also provide a signature of the mixed BEC 
state, vanishing as the boundary to the regime of phase separation is approached. 
\end{abstract}


\maketitle

\section{Introduction}
In recent years, ultracold atomic gases have emerged as a remarkable
new setting to observe novel many-body phenomena. 
Following earlier achievements, such as
 artificial gauge fields~\cite{prl:spielman2009} and  
artificial magnetic fields~\cite{nat:spielman2009} for cold atoms, recently
the Spielman group at NIST has realized light-induced 
artificial spin-orbit coupling (SOC) of a $^{87}$Rb Bose-Einstein condensate (BEC)~\cite{nat:lin}.  In such 
experiments, dressed atomic spin states with emergent SOC are engineered via 
coupling to Raman lasers.
  This experimental knob further expands the 
space of Hamiltonians for cold atom systems to realize, and opens the
possibility of simulating solid-state systems in which SOC plays a
role, including the spin Hall effect~\cite{sci:kato}, Majorana fermions~\cite{Mourik}, and  topological 
insulating phenomena~\cite{rmp:kane2010,QiZhang}.

Theoretical interest in bosons with SOC has been strong for many years,
although many early papers focused on the case of Rashba-type spin-orbit 
coupling~\cite{pra:Stanescu2008,prl:WangZhai,pra:xu,pra:kawakami,chinphys:wu,prl:HuPu,prl:sinha}.  
The NIST Raman 
setup instead realizes SOC only along one direction, i.e., the SOC Hamiltonian is
of the form $\Hh_{\rm SOC} \propto \sigma_z p_x$, with $p_x$ the momentum operator
along the $x$ direction and $\sigma_z$ the Pauli matrix acting in the space of 
dressed spins. 
  Subsequent experiments have observed 
dipole oscillations of bosons with artificial SOC~\cite{prl:pan} and studied their phases at finite 
temperature~\cite{preprint:zhang}, have realized light-induced
SOC for cold fermionic gases~\cite{prl:WangFermi,prl:CheukFermi}, 
and have employed a similar setup to observe Zitterbewegung of bosons described by an effective Dirac 
Hamiltonian~\cite{arxiv:ZhangEngelsZitter,arxiv:spielmanZitter}

A key observation of Ref.~\onlinecite{nat:lin} was the
phase transition from a mixed BEC phase, with condensates of
both dressed spin states ($|\downarrowp\rangle$ and $|\uparrowp\rangle$),
into a regime of phase separation, with spatially separated $|\downarrowp\rangle$ 
and $|\uparrowp\rangle$ condensates. 
 Here, the dressed states $|\downarrowp\rangle$ and $|\uparrowp\rangle$  
states emerge from the Raman laser coupling to two hyperfine levels ($|F,m_F\rangle$) of  $^{87}$Rb, with 
$|\!\!\uparrow\rangle = |1,0\rangle$  and $|\!\!\downarrow\rangle = |1,-1\rangle$.  Theoretically
this mixed phase is predicted to exhibit \lq\lq stripe order\rq\rq\ in the form of density
modulations along the $x$ axis, due to the system condensing in a superposition of states with
different momenta~\cite{prl:ho2011,prl:stringari,arxiv:zhai2012}, although such density modulations
may be difficult to observe.

\begin{figure}[!htb]
  \centering
\includegraphics[width=\columnwidth]{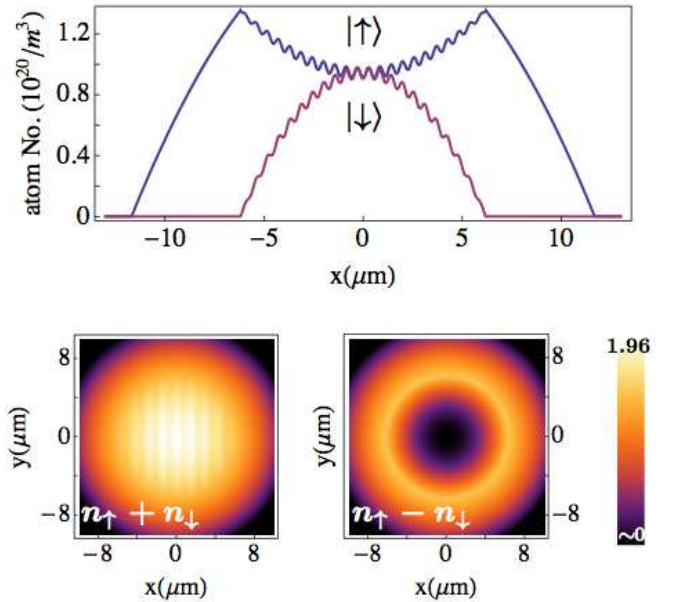}\\
 \caption{(Color online) Atom density profiles for a BEC with light-induced SOC.
The upper plot shows the densities of 
spins-$\uparrow$ and spins-$\downarrow$ at $y=z=0$ as a function of
position ($x$), showing a central core of mixed BEC and an outer
shell of $\uparrowp$ BEC and exhibiting a nonmonotonic
density profile for the $\uparrow$ atom density in 
mixed region.  The small density oscillations reflect the \lq\lq stripe\rq\rq\
order~\cite{prl:ho2011,prl:stringari,arxiv:zhai2012} in this phase.
The lower plots show a top view of the total
density, $n_\uparrow+ n_\downarrow$, and magnetization $n_\uparrow-n_\downarrow$
for the same parameters (given in the text).  The density scale in the lower plots is
atom number in $10^{20}/m^3$. }\label{fig:PrimedNOriginalDensityDis}
\end{figure}

The purpose of this paper is to predict additional experimental signatures of the mixed BEC
phase of bosons with light-induced SOC and of the transition to the phase separated state, taking into 
account the spin-dependent interactions of  $^{87}$Rb~\cite{njp:Widera} that exhibit a repulsion among
the $\uparrow$ species that is larger than the repulsion among the  $\downarrow$
species.  This asymmetry necessitates applying a negative Zeeman energy difference between 
the two species ($\delta<0$, lowering
the energy of $\uparrow$ bosons relative to the $\downarrow$ bosons) 
to stabilize the mixed BEC state~\cite{nat:lin}.  We find this further implies that a trapped gas 
stabilizing the mixed phase will generally possess an outer shell of $\uparrow'$ BEC, shown in Fig.~\ref{fig:PrimedNOriginalDensityDis}.  
This simply follows from the fact that the mixed BEC arises from atomic interactions, which are
 smaller near the cloud edge, where densities are smaller, and the system will locally
establish a $\uparrow$ BEC, since this is the lowest energy state.  

    We furthermore find an unusual density profile for the 
mixed phase in a trap: Due to the dependence of the interactions between dressed states on
the Raman coupling strength, we find the local density of $\uparrowp$ bosons {\it increases\/}
with increasing radius, in contrast to the $\downarrowp$ bosons that exhibit the conventional density profile, i.e.,
a  density that decreases with increasing radius.  This predicted density dependence follows from our
analysis of the fixed chemical potential phase diagram along with the local density approximation (LDA). 

We also study signatures of the mixed BEC phase in dynamics~\cite{prl:ZhangZhang,pra:zhengli2012,pra:chenzhai2012},
in particular focusing on the Bogoliubov sound mode of the mixed BEC phase, a well-known signature
of superfluidity that can be measured via Bragg spectroscopy~\cite{prl:Steinhauer}.  We find a collective
 Bogoliubov mode with a velocity that is suppressed with increasing light-atom coupling,
vanishing at the phase boundary to the 
regime of phase separation.

This paper is organized as follows.  
In Sec.~\ref{sec:model}, we recall the model Hamiltonian for bosons with Raman laser-induced 
SOC as realized in Ref.~\onlinecite{nat:lin} and outline the mapping to a low-energy
Hamiltonian description of the dressed spin states. 
In Sec.~\ref{pdfc}, we use the Gross-Pitaevskii equations to derive the mean-field phase diagram for this low-energy Hamiltonian at 
fixed chemical potentials for the dressed spin states, and discuss the connection to the 
phase diagram at fixed number.
In Sec.~\ref{sec:tbsoc} we employ the mean-field Gross-Pitaevskii equations along with the local density approximation 
to predict the spatial profile for the dressed spins (and for the original spin states) in a harmonic trap.  
In Sec.~\ref{sec:sm} we present our results using the time-dependent Gross-Pitaevskii equations to derive the 
Bogoliubov modes for the mixed BEC phase.  
In Sec.~\ref{sec:cr}, we provide some brief concluding remarks.  Appendix~\ref{Eq:appendix} provides
some technical details of the mapping to the low energy effective Hamiltonian.

\section{Model}
\label{sec:model}
 The setup of
Ref.~\cite{nat:lin} uses a pair of Raman lasers to
couple two atomic hyperfine Zeeman levels of $^{87}$Rb.
 In the rotating-wave approximation,
and focusing on the $m=0$  and $m=-1$ subspace (represented by the fields $\Psi_\uparrow$ and $\Psi_\downarrow$
respectively)
the single-particle Hamiltonian is $\mathcal{H}_0=\int d^3 r ~\Psi^\dagger(\br)\Hh\Psi(\br)$,
with
\bea
\Hh\equiv \left(
\begin{array}{cc}
\f{p^2}{2m}+\f{\delta}{2} & \f{1}{2}\Omega e^{2i\bk_L \cdot \br} \\
\f{1}{2}\Omega e^{-2i\bk_L \cdot \br} & \f{p^2}{2m}-\f{\delta}{2} 
\end{array}
\right),
\label{Eq:hhat}
\eea
where  $\Psi(\br) = \begin{pmatrix}\Psi_\uparrow(\br)& \Psi_\downarrow(\br)\end{pmatrix}^T$.
The diagonal terms of Eq.~(\ref{Eq:hhat}) describe the
atom kinetic energy ($\bp = -i\hbar\grad$) with mass $m$ and the Zeeman energy difference $\delta$, controlled by an external magnetic field. 
 The off-diagonal
terms capture the Raman coupling of the spin-$\uparrow$ and spin-$\downarrow$ states, parameterized by
$\Omega$ and the wavevector $\bk_L = k_L \xh$.  The spin-orbit coupling form of $\Hh$ 
emerges once we use the unitary operator $\Uh = {\rm e}^{i\bk_L \cdot \br \sigma_z}$ (with $\sigma_z$ the Pauli matrix)  to 
 rotate the Hamiltonian matrix to $\Hh_r = \Uh^\dagger \Hh \Uh$ with~\cite{nat:lin} 
\be
\label{Eq:rotated}
\Hh_r(\bp,\delta) = \frac{1}{2m}(\bp^2+\bk_L^2) + \frac{1}{2}\delta \sigma_z + \frac{1}{2}\Omega \sigma_x+  \frac{1}{m}k_L \sigma_z p_x,
\ee
with the final term being the effective light-induced spin-orbit coupling, 
$\Hh_{\rm SOC} = \frac{1}{m}k_L \sigma_z p_x$.
In Eq.~(\ref{Eq:rotated}) and below, we choose units such that $\hbar = 1$.

\begin{figure}[htb]
  \begin{center}
\includegraphics[width=0.92\columnwidth]{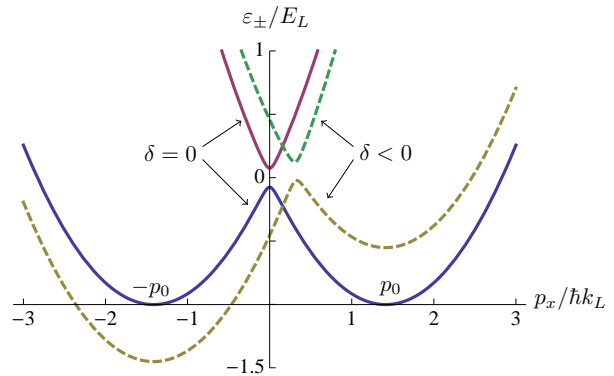}\\
  \end{center}
  \caption{(Color online) Plot of the eigenvalues $\varepsilon_\pm(\bp)$, at $p_y = p_z=0$ and
as a function of $p_x$.  The solid lines show the case of $\delta = 0$, while the 
dashed lines show the experimentally-relevant case of $\delta<0$.  The left and right minima
are the $\uparrowp$ and $\downarrowp$ dressed states, respectively.
 In the absence of interactions,
the system will condense into the left minimum (the $\uparrowp$ dressed state).  }
\label{fig:MinusBandDispersion}
\end{figure}

After making this unitary rotation, it is straightforward
 to obtain the eigenvalues of $\Hh_r(\bp,\delta)$:
\be
\label{Eq:epspm}
\varepsilon_\pm(\bp)=\frac{p^2+k_L^2}{2m} \pm \sqrt{\frac{\Omega^2+\delta^2}{4} + \frac{k_L^2 p_x^2}{m^2} +\frac{\delta k_L p_x}{m}},
\ee
plotted in Fig.~\ref{fig:MinusBandDispersion} for the 
case of $\delta = 0$ (solid curves) and 
 $\delta>0$ (dashed curves).  
Here, we're mainly interested in the regime in which the 
lower band $\varepsilon_-(\bp)$
possesses the double-well shape shown in the figure (for sufficiently small $\Omega$ and $\delta$).
Following Lin et al~\cite{nat:lin}, we proceed to construct a low-energy Hamiltonian focusing on states near these two minima
 (occuring at $\pm p_0$ with $p_0\approx k_L$).  With the details relegated to the Appendix~\ref{Eq:appendix}, we find the
approximate form of the single-particle Hamiltonian:
\be
 \curH_0 = \sum_{\bp}\sum_{\sigma= \uparrowp,\downarrowp} 
(\varepsilon_\sigma(\bp) -\mu_\sigma\big)\psi_\sigma^\dagger(\bp)\psi_\sigma^\phdag(\bp),
\label{Eq:simplify2}
\ee
 where we included a chemical potential $\mu$ that couples to the density and defined $\mu_\uparrowp = \mu -\frac{1}{2}\delta$
and $\mu_\downarrowp = \mu +\frac{1}{2}\delta$.  Here,  $\psi_\sigma^\phdag(\bp)$ is an annihilation operator for a bosonic dressed spin state,
and the effective dispersion is 
\be
\label{Eq:effectivedispersion}
\varepsilon(\bp) =\frac{1}{2m^*} p_x^2 + \frac{1}{2m}(p_y^2+p_z^2),
\ee
equal to the bare dispersion in the $y$ and $z$ directions, and reflecting the curvature of the minima of $\varepsilon_-(\bp)$, 
that satisfies $(m^*)^{-1} = m^{-1} (1-\Omegah^2)$, in the $x$ direction.  The dimensionless coupling $\Omegah \equiv \Omega/4E_L$ with
$E_L \equiv k_L^2/2m$.  

As discussed in Appendix~\ref{Eq:appendix}, Eq.~(\ref{Eq:simplify2}) is valid at sufficiently small atom-light 
coupling and Zeeman
energy difference, i.e., $\Omegah\ll 1$ and $\deltah \ll 1$, where $\deltah = \delta/4E_L$ is the corresponding dimensionless
Zeeman energy difference.  Within a similar approximation scheme, the interaction Hamiltonian for the dressed spins is:
\bea
&&\curH_1 = \frac{1}{2}\int d^3r \Big[ 
g_{\uparrowp\uparrowp}
|\psi_\uparrowp(\br)|^4+ g_{\downarrowp\downarrowp}|\psi_\downarrowp(\br)|^4
\nonumber \\
&&\quad \quad +
2 g_{\uparrowp\downarrowp}|\psi_\uparrowp(\br)|^2|\psi_\downarrowp(\br)|^2\Big],
\label{eq:lowenergyhone}
\eea
where $\psi_\sigma(\br)$ is the corresponding field operator, the Fourier
transform of $\psi_\sigma(\bp)$.  The interaction parameters are~\cite{nat:lin}:
\bea
g_{\uparrowp\uparrowp} &=& c_0,
\\
 g_{\downarrowp\downarrowp} &=& c_0 + c_2,
\\
g_{\uparrowp\downarrowp} &=&  c_0(1+\Omegah^2)+c_2,
\eea
with the couplings ~\cite{njp:Widera} $c_0 = 4\pi (a_0+2a_2)/3m$ and $c_2 = 4\pi (a_2-a_0)/3m$.
For $^{87}$Rb, the scattering lengths $a_2$ and $a_0$ are almost equal (with
$a_2- a_0\simeq -1.07\abohr $ with $\abohr$ the Bohr radius), implying $c_2<0$ (inducing mixing among the
two spin-states) and $|c_2|\ll c_0$.

\begin{figure}[!htb]
  \begin{center}
\includegraphics[width=0.89\columnwidth]{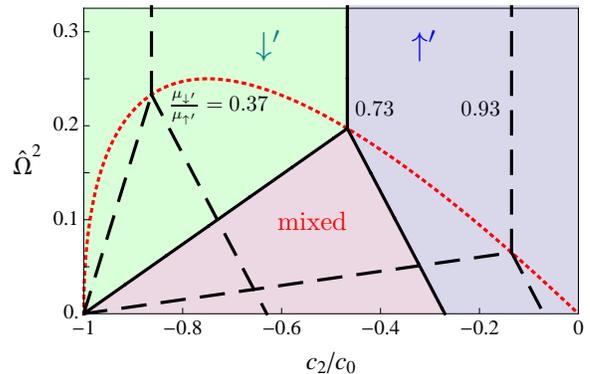}\\
  \end{center}
\vspace{-0.5cm}
  \caption{(Color online) The solid lines show the phase diagram, at fixed $\mu_\downarrowp/\mu_\uparrowp = 0.73$,
separating regions of BEC $\downarrowp$ (upper left, green online), BEC $\uparrowp$ (upper right, blue online),
and a mixed BEC of both species (lower triangle, red online).  The two sets of dashed lines show the 
phase diagram at two additional values of $\mu_\downarrowp/\mu_\uparrowp$, showing the evolution of 
the phase diagram as a function of this ratio.
For experiments at fixed particle number, the relevant phase boundary is the dotted line: Below this dotted line, the mixed BEC is stable,
while above this dotted line the system will phase separate into regions of uniform $\uparrowp$ superfluid and uniform $\downarrowp$ superfluid.}
\label{fig:CombinedPhaseDiag}
\end{figure}

\section{Phase diagram at fixed chemical potential}
\label{pdfc}
In the present section, we analyze the phase diagram at fixed chemical potentials for the two
species, using
the effective low-energy Hamiltonian $\curH = \curH_0 + \curH_1$ given by Eqs.~(\ref{Eq:simplify2})
and (\ref{eq:lowenergyhone}) of the preceding section.  In the spirit of mean-field
theory, we assume spatially-uniform expectation values,  $\langle \psi_{\sigma'}\rangle$, for
each species, and minimize the grand free energy.   We find four distinct solutions: The 
trivial noncondensed solution $\langle \psi_{\uparrowp}\rangle = \langle \psi_{\downarrowp}\rangle = 0$, 
and  solutions in which one or both of $\psi_\uparrowp$ or $\psi_\downarrowp$ is condensed.
The latter follow from the Gross-Pitaevskii (GP) equations (where we henceforth drop the angle
brackets on $\psi_\uparrowp$ and $\psi_\downarrowp$ for simplicity):
%
%
%
\bse
\label{Eq:els}
\bea
\mu_\uparrowp &=&  c_0 |\psi_\uparrowp|^2  + 
\big[ c_0(1+\Omegah^2)+c_2  \big]
|\psi_\downarrowp|^2 ,~~
\\
\mu_\downarrowp  &=&   (c_0+c_2) |\psi_\downarrowp|^2  + 
 \big[ c_0(1+\Omegah^2)+c_2  \big]
|\psi_\uparrowp|^2 ,~~
\eea
\ese
which exhibit three solutions.  Two of these solutions refer
to the case in which only one of $\psi_\downarrowp$ or $\psi_\uparrowp$ is
condensed:
\bse
\label{eq:udsol}
\bea
\label{downsolution}
&&\psi_\uparrowp  = 0;\hspace{1cm} n_\downarrowp = \frac{\mu_\downarrowp}{c_0+c_2}, 
\\
&&\psi_\downarrowp  = 0;\hspace{1cm} n_\uparrowp = \frac{\mu_\uparrowp}{c_0},
\label{upsolution}
\eea
\ese
that we call the BEC $\downarrowp$ and BEC $\uparrowp$ phases (or, 
simply, $\downarrowp$ and $\uparrowp$), respectively,  referring to the
condensed species.  Here, we introduced the notation $n_{\sigma} = |\psi_{\sigma}|^2$ for the mean-field 
densities of the two species.  The last solution is the mixed phase, in which both species are condensed.  Solving
Eqs.~(\ref{Eq:els}) for $n_\uparrowp$ and $n_\downarrowp$ gives:
\be
\begin{pmatrix} 
n_\uparrowp \\
n_\downarrowp \end{pmatrix}  = \frac{1}{D}
\begin{pmatrix} c_0 + c_2 & - c_0(1+\Omegah^2)-c_2  \\
- c_0(1+\Omegah^2)-c_2   & c_0 \end{pmatrix}
   \begin{pmatrix} \mu_\uparrowp \\ \mu_\downarrowp \end{pmatrix} ,
\label{eq:mixedsolution}
\ee
where we defined the denominator 
\be
\label{Eq:denominator}
D =(c_0+c_2)c_0 - (c_0(1+\Omegah^2)+c_2)^2,
\ee
%
which, in the fixed-number ensemble, determines the phase-separation boundary which is $D=0$ (as discussed below).

In the case of positive chemical potentials for the two species, the abovementioned 
trivial solution $\psi_\uparrowp = \psi_\downarrowp=0$
 never occurs (note we focus on the zero temperature case $T=0$), and, for
any given chemical potential ratio $\mu_\downarrowp/\mu_\uparrowp$, 
the phase diagram exhibits the three remaining phases: $\uparrowp$, $\downarrowp$, and
mixed.  For the case of $\mu_\uparrowp = \mu_\downarrowp$, the mixed phase is never stable,
and the system always exhibits the $\downarrowp$ phase.  This can be traced to the fact
that, as noted above, the interaction Hamiltonian is intrinsically \lq\lq imbalanced\rq\rq,
favoring the $\downarrowp$ state, so that a nonzero chemical potential imbalance $\delta <0$, or $\mu_\downarrowp<\mu_\uparrowp$
is needed to attain the mixed phase. 
Thus, henceforth we focus on the regime of $\delta<0$.

The solid lines in Fig.~\ref{fig:CombinedPhaseDiag} show the ground-state phase diagram in the fixed chemical potential
ensemble, at $\mu_\downarrowp/\mu_\uparrowp= 0.73$, showing regimes of $\downarrowp$ superfluid (upper-left, green),
$\uparrowp$ superfluid (upper right, blue) and mixed superfluid  (bottom center, red) phases, obtained by
directly finding the state with the lowest value of the expectation value of the free energy.
  Thus, the mixed phase is stable in a triangular region of the phase diagram,
exhibiting continuous phase transitions, with increasing normalized light-atom coupling $\Omegah$, to the 
$\downarrowp$ superfluid (for large $|c_2|/c_0$, to the left in the phase diagram) and to the  
$\uparrowp$ superfluid (for small $|c_2|/c_0$, to the right in the phase diagram).  The same structure of the 
phase diagram holds for any ratio $\mu_\downarrowp/\mu_\uparrowp$, with the three curves that separate the phases 
moving as a function of the chemical potential ratio $\mu_\downarrowp/\mu_\uparrowp$; the two sets of dashed lines
in  Fig.~\ref{fig:CombinedPhaseDiag} indicate the locations of these boundaries for $\mu_\downarrowp/\mu_\uparrowp = 0.37$
and $\mu_\downarrowp/\mu_\uparrowp = 0.93$.

At  large $\Omegah$, where the mixed phase is not stable, the phase boundary separating the $\uparrowp$ and $\downarrowp$ 
is defined by when the mean-field energies of the $\uparrowp$ and $\downarrowp$ are equal.  Since 
the expectation value of 
$\curH_1$, Eq.~(\ref{eq:lowenergyhone}), is independent of $\Omegah$
in the $\uparrowp$ and $\downarrowp$ phases (because $\Omegah$ only enters the final term
of Eq.~(\ref{eq:lowenergyhone}), which vanishes in this phase),
this boundary must be independent of $\Omegah$, i.e. vertical in Fig.~\ref{fig:CombinedPhaseDiag}.  Equating
these energies gives 
\be
\label{eq:verticalcrossing}
c_2 = c_0\Big(\frac{\mu_\downarrowp^2}{\mu_\uparrowp^2} -1 \Big),
\ee
for the critical coupling separating these phases.  

At low values of $\Omegah^2$, the mixed phase
 is stable for a range of $c_2$ values as shown in Fig.~\ref{fig:CombinedPhaseDiag},
and exhibits condensate densities in the $\uparrowp$ and $\downarrowp$ states described
by Eq.~(\ref{eq:mixedsolution}).  The transition out of the mixed phase
occurs when, with increasing $\Omegah^2$, one of $n_\uparrowp$ or $n_\downarrowp$ vanishes, 
leaving a condensate of the other species.  Thus, the phase boundary for
the mixed-$\downarrowp$ transition occurs when $n_\uparrowp \to 0$ in Eq.~(\ref{eq:mixedsolution}):
\be
\label{downboundary}
\Omegah^2 = \big( 1+ \frac{c_2}{c_0}\big) \big( \frac{\mu_\uparrowp}{\mu_\downarrowp} - 1\big),
\ee
while the phase boundary for the mixed-$\uparrowp$ transition,
\be
\label{upboundary}
\Omegah^2 = - \frac{c_2}{c_0} + \big( \frac{\mu_\downarrowp}{\mu_\uparrowp} - 1\big),
\ee
occurs when $n_\downarrowp \to 0$.  The three curves Eq.~(\ref{eq:verticalcrossing}),
Eq.~(\ref{downboundary}), and Eq.~(\ref{upboundary}) thus determine the fixed
chemical potential phase diagram.

The dotted red line in this figure  Fig.~\ref{fig:CombinedPhaseDiag}, determined by the vanishing of Eq.~(\ref{Eq:denominator}), i.e., $D=0$, 
 shows how the intersection of the phase boundaries evolves as
a function of $\mu_\downarrowp/\mu_\uparrowp$.  However, it also indicates the phase boundary for the SOC 
boson gas at {\em fixed density\/}, 
with the mixed BEC phase stable for $D>0$ and unstable to phase separation for $D<0$.
 To see this, note that 
the mixed phase at fixed particle numbers $N_\downarrowp$ and $N_\uparrowp$ (or fixed $N_\downarrow$ and $N_\uparrow$)
 can be regarded as having resulted from a system at fixed $\mu_\uparrowp$
and $\mu_\downarrowp$ with the chemical potentials adjusted to satisfy the fixed-number requirement.
Starting from the mixed phase, as $\Omegah$ is adjusted upwards towards the red dotted line,  $\mu_\uparrowp$
and $\mu_\downarrowp$ will adjust to maintain the imposed values of $N_\downarrowp$ and $N_\uparrowp$.  
However, beyond the red dotted line, it is no longer possible for the chemical potentials to adjust
to attain a stable mixed phase, and the system phase separates into uniform BEC $\uparrowp$ and BEC $\downarrowp$ 
to satisfy the fixed-number constraint. 

The same result for the boundary separating the mixed BEC and phase-separation regimes can be found
by directly computing the expectation value of the Hamiltonian, at fixed particle number, assuming
either a homogeneous mixed phase or a phase separated BEC and equating the energies, as found by Lin et al~\cite{nat:lin}.
Before proceeding, we note that our result for the phase diagram at fixed chemical potentials agrees,
in the case of 
$\mu_\uparrowp = \mu_\downarrowp$, with the results of Ho and Zhang (i.e., Fig.3 of Ref~\cite{prl:ho2011}),  although our axes
and notation are different.  The evolution of this phase diagram as a function of $\mu_\uparrowp$ and $\mu_\downarrowp$, 
will be essential to study the case of a trapped BEC with SOC, discussed in the next section.

\section{Trapped bosons with SOC}
\label{sec:tbsoc}
In the preceding section, we determined the phase diagram for a uniform boson gas with artificial light-induced SOC 
in the ensemble of fixed 
chemical potentials $\mu_\uparrowp$ and $\mu_\downarrowp$, showing how it can be used to obtain the boundary to the
regime of phase separation
in the fixed number ensemble.  In the present section, we 
 turn to the question of the density 
distribution of the two boson species in a parabolic (harmonic) trap, making use of the fixed $\mu_\uparrowp$ and $\mu_\downarrowp$
results of the preceding section.

We consider an anisotropic trapping geometry, 
\be
\vtrap(\br)=\frac{1}{2}m(\Omega_z^2 z^2+\Omega_s^2 s^2),
\ee
 where $s^2=x^2+y^2$.  Below, we'll make the choice $\Omega_z>\Omega_s$ for the trapping frequencies, such that an oblate ``pancake\rq\rq\ cloud shape is 
expected.  Our analysis of the density distributions in the presence of the trap uses  the local density approximation (LDA).  Within the LDA,
the densities $|\psi_{\uparrow^\prime}|^2$ and $|\psi_{\downarrow^\prime}|^2$ are given by the uniform-case results 
Eq.~(\ref{eq:mixedsolution}) but with $\mu_\sigma \to \mu_\sigma - \vtrap(\br)$
(where now $\mu_\sigma$ is the chemical potential at the trap center,  $r=0$).  After some simplification, these densities can be written as
\bse
\label{denlda}
\bea
\label{uparrowplda}
\abs{\psi_{\uparrow^\prime}}^2 &= \f{\tilde{\mu}_{\uparrow^\prime}}{\tilde{g}_{\uparrow^\prime}} (1+\f{z^2}{{\tilde{R}_{z{\uparrow^\prime}}}^2}+\f{s^2}{{\tilde{R}_{s{\uparrow^\prime}}}^2}),\\
\abs{\psi_{\downarrow^\prime}}^2 &= \f{\tilde{\mu}_{\downarrow^\prime}}{\tilde{g}_{\downarrow^\prime}} (1-\f{z^2}{{\tilde{R}_{z{\downarrow^\prime}}}^2}-\f{s^2}{{\tilde{R}_{s{\downarrow^\prime}}}^2}),
\label{downarrowplda}
\eea
\ese
where we defined effective interaction parameters $\tilde{g}_{\uparrowp} =-D/c_0\Omegah^2$  and 
$\tilde{g}_{\downarrowp} =-D/(c_2 + c_0\Omegah^2)$, 
with  $D$ defined in Eq.~(\ref{Eq:denominator}) above, and the effective
chemical potentials
\bea 
\tilde{\mu}_{\uparrow^\prime}&=&\frac{(c_0(1+\Omegah^2)+c_2)\mu_{\downarrow^\prime}-(c_0+c_2)\mu_{\uparrow^\prime}}{c_0 
{\hat{\Omega}}^2} ,
\\
\tilde{\mu}_{\downarrow^\prime}&=&\frac{(c_0(1+\Omegah^2)+c_2)\mu_{\uparrow^\prime}-c_0\mu_{\downarrow^\prime}}{c_2+c_0 
{\hat{\Omega}}^2} ,
\eea 
where, crucially, the ratios $\tilde{\mu}_{\sigma}/\tilde{g}_{\sigma}>0$ for both $\uparrowp$ and $\downarrowp$, so that the densities in 
Eq.~(\ref{denlda}) are positive.  The radii
 $\tilde{R}_{s\sigma}$ and $\tilde{R}_{z\sigma}$, which determine the spatial variation of the densities in the plane of the 
 pancake shaped cloud and perpendicular to it, respectively,
are given by 
\bea
\tilde{R}_{z{\uparrow^\prime}}= \sqrt{\f{-2\tilde{\mu}_{\uparrow^\prime}}{m {\Omega_z}^2}},~\tilde{R}_{s{\uparrow^\prime}}=\sqrt{\f{-2\tilde{\mu}_{\uparrow^\prime}}{m {\Omega_s}^2}},\\
\tilde{R}_{z{\downarrow^\prime}}= \sqrt{\f{2\tilde{\mu}_{\downarrow^\prime}}{m {\Omega_z}^2}},~\tilde{R}_{s{\downarrow^\prime}}=\sqrt{\f{2\tilde{\mu}_{\downarrow^\prime}}{m {\Omega_s}^2}},
\eea
Although Eqs.~(\ref{denlda}) are similar to the usual LDA form for the density variation of a trapped BEC, one unusual feature stands out:
While $\abs{\psi_{\downarrow^\prime}}^2$ decreases with increasing radius, the $\uparrow'$ density {\em increases\/} with increasing radius.
This behavior only occurs in the mixed phase which, for typical experimentally-relevant parameters, will occur in the trap center.  For further
increasing radius, $\abs{\psi_{\downarrow^\prime}}^2 \to 0$ in the usual Thomas-Fermi fashion and beyond this radius the system 
is locally in a BEC of the spins-$\uparrowp$.

In Fig.~\ref{fig:PrimedNOriginalDensityDis}, we show the actual bosons densities  $\abs{\Psi_{\uparrow}}^2$ and $\abs{\Psi_{\downarrow}}^2$,
that are related to  $\abs{\psi_{\uparrowp}}^2$ and $\abs{\psi_{\downarrowp}}^2$ via
\bea
\label{equ:RelatingUpDensities}
&&\hspace{-.75cm}|\Psi_\uparrow(\br)|^2  = |\psi_\uparrowp(\br) -
\frac{1}{2}\Omegah {\rm e}^{2ik_L x} \psi_\downarrowp(\br)|^2,
\\
&&\hspace{-.75cm}|\Psi_\downarrow(\br)|^2 = 
 |\psi_\downarrowp(\br) - \frac{1}{2}\Omegah {\rm e}^{-2ik_L x} \psi_\uparrowp(\br)|^2 ,
\label{equ:RelatingDownDensities}
\eea
which follow from Eq.~(\ref{Eq:psiup}) in the limit of small $\Omegah$ and $\deltah$.  
In Eqs.~(\ref{equ:RelatingUpDensities}) and (\ref{equ:RelatingDownDensities}), we take $\psi_\uparrowp(\br)$
and
 $\psi_\downarrowp(\br)$
to be real and positive. The relative phase between these condensates, yielding
the minus signs in these expressions, follows by assuming the system will minimize
the interaction energy density (and therefore $|\Psi_\uparrow(\br)|^2$ and $|\Psi_\downarrow(\br)|^2$)
at the trap center.

Note that, since $\Omegah\ll 1$ to stabilize
the mixed phase, the density  $n_\sigma(\br) = |\Psi_\sigma(\br)|^2$ is approximately equal to the corresponding primed density plus an 
$\curO(\Omegah)$ term (the cross term upon expanding the modulus squared), leading to a $\cos 2k_Lx$ spatial modulation
(or, stripe order~\cite{prl:WangZhai}).  This oscillatory spatial variation is, however, only barely visible in 
 Fig.~\ref{fig:PrimedNOriginalDensityDis} in the central mixed-BEC region, due to the smallness of $\Omegah$.

In Fig.~\ref{fig:PrimedNOriginalDensityDis}, we chose parameters 
consistent with those of Ref.~\onlinecite{nat:lin}: 
Trapping frequencies $\Omega_s=2\pi\times 50$ Hz, $\Omega_z=2\pi\times 140$ Hz, interaction parameters
$c_0=h\times7.79\times 10^{-12}$ Hz cm$^3$, $c_2=-h\times 3.61\times 10^{-14}$ Hz cm$^3$, the wavevector $k_L = \sqrt{2}\pi/804.1$nm, and the
spin-orbit coupling parameter $\Omega=0.15 E_R$.  The chemical potentials $\mu_\downarrowp = 1464$Hz  and $\mu_\uparrowp = 1467$Hz were chosen to 
achieve a total particle number $N= 180,000$ and reflect an effective Zeeman 
field $|\delta| = |\mu_\uparrowp - \mu_\downarrowp| = 3$Hz (also consistent with Ref.~\onlinecite{nat:lin}).

Next we present a physical picture of the density profile results.  The sequence of phases, within the LDA, in fact follows directly
from the structure of the fixed-$\mu$ phase diagram.  To see this, we note that, as seen in Fig.~\ref{fig:CombinedPhaseDiag},
the \lq\lq triangle\rq\rq\ of stable mixed phase moves to the left with decreasing $\mu_\downarrowp/\mu_\uparrowp$, with
the $\downarrowp$ condensate always occuring to the left of this triangle and the $\uparrowp$ condensate always occuring
to the right.  Within the LDA, then, the quantity to consider is the spatially-varying effective chemical potential
ratio $\gamma(\br)\equiv [\mu_{\downarrow^\prime}-\vtrap(\br)]/[\mu_{\uparrow^\prime}-\vtrap(\br)]$, which 
decreases with increasing $\br$ (when $\mu_\downarrowp<\mu_\uparrowp$, which is required for stability of
the mixed phase).  If the mixed phase is stable in the center, then
this implies that, at $\br = 0$, the system parameters must put it in the triangle of mixed BEC phase of 
Fig.~\ref{fig:CombinedPhaseDiag}.   Increasing radius will decrease $\gamma(\br)$, moving the triangle
of mixed BEC phase to the left, leaving the system locally in the $\uparrowp$ phase at the edge. 
Another logical possibility, 
in which the $\downarrowp$ phase is stable in the center, followed by the mixed phase at 
intermediate radii, followed by the $\uparrowp$ phase at large radii, is possible but turns out to be difficult to 
achieve using experimentally-realistic parameters.  

The outer shell of $\uparrowp$ condensate is described by the standard local density approximation
for a single-species BEC, with $\abs{\psi_{\uparrow^\prime}(\br)}^2=(\mu_{\uparrow^\prime}-\vtrap(\br))/c_0$.
As we have already mentioned, the existence of the outer shell of $\uparrowp$ BEC is generally expected,
since the mixed phase is stabilized by interactions.  At large radii, where the atom densities are small, 
interactions can be neglected, and the system condenses into the lowest state, i.e., the left
minimum of Fig.~\ref{fig:MinusBandDispersion}, which is the $\uparrowp$ phase.  Therefore, we generally
expect the outer shell of $\uparrowp$ condensate.  With decreasing radius, coming in from the outside
of the cloud, interaction effects eventually favor the population of the right minimum of 
Fig.~\ref{fig:MinusBandDispersion}, so that the system locally enters the mixed phase.

To understand the behavior of the densities in the central mixed BEC region, we 
transform the interaction Hamiltonian Eq.~(\ref{eq:lowenergyhone}) to the basis of magnetization
($M = n_\uparrowp-n_\downarrowp$) and total density ($n=  n_\uparrowp+n_\downarrowp$) :
\bea
&&\curH_1 = \frac{1}{2}\int d^3r \Big[  \big(c_0 + \frac{1}{2}c_0 \Omegah^2 + \frac{3}{4}c_2\big)n^2(\br)
\\
\nonumber &&
- \big(  \frac{1}{4}c_2 + \frac{1}{2}c_0\Omegah^2\big)M^2(\br) - \frac{1}{2}c_2 M(\br)n(\br) \Big].
\eea
Recall that $c_0\gg |c_2|$ and $\Omegah^2 \ll 1$.  This implies that, in the first term,
the overall density is controlled by $c_0>0$, so that $n(\br)$ should
exhibit the standard parabolic Thomas-Fermi profile in a trap.  The magnetization $M(\br)$, however, does not 
directly couple to the trap potential, but exhibits a spatial variation since the last term couples
$M(\br)$ and $n(\br)$.   Since $c_2<0$, this term favors having small (or negative) $M(\br)$ 
in region of large $n(\br)$ (i.e., at the trap center), leading to the
 central dip in the magnetization shown in the right lower panel of Fig.~\ref{fig:PrimedNOriginalDensityDis}.

\section{Sound Mode}
\label{sec:sm}
 In the preceding section, we showed that the mixed BEC phase of 
 bosons with SOC exhibits an unusual density profile for the two species in a harmonic
trapping potential.  Now we turn to another signature of the mixed BEC phase, which is the 
Bogoliubov sound velocity, focusing on the case of a uniform condensate.

 Using the effective Hamiltonian for the $\uparrowp$ and $\downarrowp$
states, consisting of Eq.~(\ref{Eq:simplify2}) and Eq.~(\ref{eq:lowenergyhone}), we have the 
time-dependent GP equations (recall $\hbar = 1$):
\bea
&&\hspace{-0.75cm}(i  \partial_t\!- \!\varepsilon(\bp)+\mu_\uparrowp)\psi_\uparrowp\! =  \!  c_0 |\psi_\uparrowp|^2 \psi_\uparrowp   + 
\bar{c}
|\psi_\downarrowp|^2  \psi_\uparrowp 
\\
&&\hspace{-0.75cm}(i  \partial_t\! - \!\varepsilon(\bp)+\mu_\downarrowp)\psi_\downarrowp 
\!=\!  (c_0+c_2) |\psi_\downarrowp|^2 \psi_\downarrowp  + 
\bar{c}
|\psi_\uparrowp|^2 \psi_\downarrowp ,
\nonumber
\eea 
where we defined $\bar{c} \equiv  c_0(1+\Omegah^2)+c_2$.  Here, $\varepsilon(\bp)$ is the 
 effective dispersion Eq.~(\ref{Eq:effectivedispersion}), and $\bp = -i \grad$ is
the momentum operator.

 The next step is to consider small time-dependent fluctuations $\phi_\sigma(\br,t)$ around
the equilibrium mixed phase solution, writing $\psi_{\sigma}(\br,t) = \psi_{\sigma} + \phi_{\sigma}(\br,t)$,
where $\psi_{\sigma}$ is the homogeneous mixed-phase solution satisfying Eq.~(\ref{Eq:els}), that we'll take to be real below.  We can further
express the fluctuation part as 
\be
\phi_{\sigma}=u_{\sigma}(\br) e^{-i\omega t}+v_{\sigma}^* (\br) e^{i\omega t}.
\ee
Plugging this into the time-dependent GP equations, keeping only linear terms in the fluctuations, and eliminating
the chemical potentials using Eq.~(\ref{Eq:els}), we obtain 
\begin{widetext}
\be
  \begin{pmatrix}
\varepsilon(\bp) + c_0\psi_\uparrowp^2 & c_0 \psi_\uparrowp^2 & \bar{c} \psi_\uparrowp \psi_\downarrowp & \bar{c} \psi_\uparrowp \psi_\downarrowp 
\\
 -c_0 \psi_\uparrowp^2  & -\varepsilon(\bp) - c_0\psi_\uparrowp^2& -\bar{c} \psi_\uparrowp \psi_\downarrowp & -\bar{c} \psi_\uparrowp \psi_\downarrowp 
\\
 \bar{c} \psi_\uparrowp \psi_\downarrowp & \bar{c} \psi_\uparrowp \psi_\downarrowp & \varepsilon(\bp) + (c_0+c_2)\psi_\downarrowp^2 & (c_0+c_2) \psi_\downarrowp^2 
\\
  -\bar{c} \psi_\uparrowp \psi_\downarrowp & -\bar{c} \psi_\uparrowp \psi_\downarrowp &-(c_0+c_2) \psi_\downarrowp^2  & -\varepsilon(\bp) - (c_0+c_2)\psi_\downarrowp^2
\end{pmatrix} \begin{pmatrix} u_\uparrowp(\br)\\v_\uparrowp(\br) \\ u_\downarrowp(\br)\\v_\downarrowp(\br)\end{pmatrix}
= \omega \begin{pmatrix} u_\uparrowp(\br)\\v_\uparrowp(\br) \\ u_\downarrowp(\br)\\v_\downarrowp(\br)\end{pmatrix},
\ee
\end{widetext}
describing the collective Bogoliubov modes in the mixed BEC phase.  The four eigenfrequencies $\omega(\bp)$ are straightforwardly found, after assuming plane
wave solutions $u_\sigma(\br) = u_\sigma{\rm e}^{i\bp\cdot \br}$ and $v_\sigma(\br) = v_\sigma{\rm e}^{i\bp\cdot \br}$.
They are $\pm \omega_{\alpha}$ with $\alpha = \pm$ and  
\be
\omega_\pm  = \sqrt{\varepsilon(\bp)^2+\varepsilon(\bp)(A\pm \sqrt{A^2-4 D n_\uparrowp n_\downarrowp})},
\ee
where we defined 
\bea
A={c_2} {n_{\downarrow^\prime}}+{c_0} ({n_{\uparrow^\prime}}+{n_{\downarrow^\prime}}),
\eea
where $D$ is the denominator Eq.~(\ref{Eq:denominator}) that also determines the phase boundary at fixed densities, with
stability of the mixed-BEC requiring $D>0$.

\begin{figure}[htb]
  \begin{center}
\includegraphics[width=0.85\columnwidth]{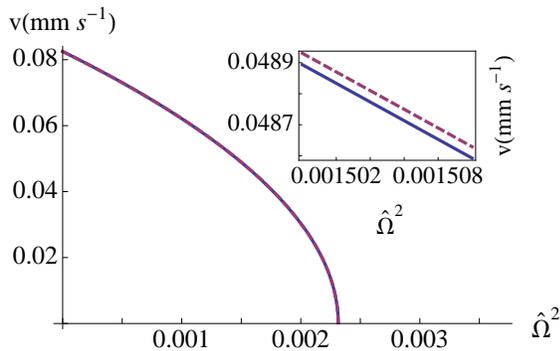}\\
  \end{center}
  \caption{(Color online) The main plot shows the Bogoliubov sound velocity in the mixed BEC phase, as a function of
normalized light-atom coupling, which vanishes at the transition to the regime of phase 
separation. At this scale, it is not possible to discern the difference between $v_x$ and
$v_\perp$ (for sound modes along the SOC direction and perpendicular to it, respectively),
although the inset, a zoom-in to these curves, shows the slight difference.  In this inset,
the dashed curve is $v_\perp$, and the solid curve is $v_x$.}
\label{fig:SoundSpeedCudVary}
\end{figure}

Although both of $\omega_\pm(\bp)$ are linearly dispersing at low $\bp$, representing Bogoliubov sound
modes for the SOC BEC, we now focus on $\omega_-(\bp)$ which has interesting behavior as a function of
the light-atom coupling.  
We first note that, due to the anisotropy of $\varepsilon(\bp)$, the corresponding sound velocity 
is smaller for modes propagating along the light-induced 
SOC direction (i.e. the $\xh$ axis) than for modes propagating perpendicular to it.  Explicitly,
 we find $v_x = v_\perp\sqrt{1-\Omegah^2}$, so that $v_x=v_\perp$ for $\Omegah = 0$ (in the limit of no light-atom coupling). 
 To obtain
$v_\perp$, we choose $\bp$ along the $\yh$ or $\zh$ direction.  Then, $v_\perp =  \d{\omega_-}{p}\mid_{p\rightarrow 0}$ with 
\be
v_\perp=
\f{1}{\sqrt{2m}}\sqrt{A-\sqrt{A^2-4 D n_\uparrowp n_\downarrowp}}.
\label{Eq:vperp}
\ee
For a spin-orbit coupled BEC in the mixed phase with fixed densities $n_\uparrowp$ and $n_\downarrowp$ (or fixed $n_\uparrow$ and 
$n_\downarrow$), Eq.~(\ref{Eq:vperp}) describes a collective superfluid sound mode.  From the form of this equation, it is
clear that $D>0$ is required and that $v_\perp\to 0$ for $D\to 0$, with increasing light-atom coupling $\Omegah$, as the 
system approaches the regime of phase separation.

 In Fig.~\ref{fig:SoundSpeedCudVary}, we illustrate this for the case of a mixed BEC state with 
$n_\uparrow = 0.6\times 10^{20}/m^3$ and $n_\downarrow = 1.3\times 10^{20}/m^3$ (with $c_0$ and $c_2$ the same as in
the preceding section).
 Note that the smallness of $c_2$ for $^{87}$Rb implies that
the mixed BEC phase is only stable for very small values of $\Omegah$, further implying that, in practice,
$v_\perp$ and $v_x$ are nearly identical for realistic parameters.  Thus, at this scale, the main plot
could be either $v_\perp$ or $v_x$.  Although the difference between these two velocities is likely not observable,
their vanishing as the phase boundary is approached would provide a distinct signature of the mixed-BEC phase. 

\section{Concluding remarks}
\label{sec:cr}
In this paper, we employed the mean-field approximation to study a $^{87}$Rb BEC with light-induced artificial SOC following
the original setup of the Spielman group at NIST~\cite{nat:lin}.  Although previous theoretical works often
made simplifying assumptions when studying this system, such as focusing on the balanced case (i.e., Zeeman energy difference $\delta = 0$) 
or neglecting the spin-dependence of the interactions, we found that accounting for these effects leads to novel insight 
into the behavior of BEC's with artificial SOC.  

In particular, we analyzed the mean-field phase diagram as a function of $\delta$ (which is equivalent to a chemical potential
difference for the two dressed states), the Raman coupling strength $\Omega$,
and interaction parameters.  We argued that the evolution of this phase diagram as a function of chemical potentials implies 
(within the local density approximation) an unusual density dependence in a harmonic trap, with the dressed spin-$\uparrow$ ($m=0$)
bosons showing a density maximum with increasing radius, where the dressed spin-$\downarrow$ ($m=-1$) density vanishes.

Our results show that, in equilibrium, attaining the mixed phase in
a trapped BEC with SOC necessitates a population imbalance or negative
detuning $\delta$, as seen in Fig.~\ref{fig:PrimedNOriginalDensityDis},
which clearly has $N_\uparrow>N_\downarrow$,
in contrast to, e.g., Fig. 2c of Lin et al showing an approximately equal number of
the two spin states. 
  We believe this discrepancy follows
from the fact that the Lin et al experiments are not fully in spin equilibrium,
and exhibit a metastable spin-mixed phase within the "metastable window" of
Fig. 2 of Ref.~\onlinecite{nat:lin}. According to our results, in {\em equilibrium\/}, a trapped BEC with SOC must have
an overall spin imbalance and will exhibit a density profile of the form shown in Fig.~\ref{fig:PrimedNOriginalDensityDis}.  

  We also predicted that the mixed-BEC phase of bosons with artificial SOC should exhibit a Bogoliubov sound mode, the velocity of which
vanishes as the regime of phase separation is approached. 
This prediction was for the case of a uniform BEC with SOC; however, most
cold atom experiments involve a harmonically trapped atomic gas with a nonuniform
atom density.  Near the trap
center, where the atom density is nearly uniform, our calculations can approximately
apply.  Additionally, a trapped {\em uniform\/} BEC (that is
confined to a \lq\lq box\rq\rq-shaped trap) 
has been recently achieved experimentally~\cite{prl:gaunt}.

We conclude by noting a few natural extensions of our work. 
 The first such extension would be to generalize our analysis to finite temperatures and 
to larger values of the Raman parameter $\Omega$ (where the double-well structure of the dispersion vanishes~\cite{nat:lin}).
Additionally, we would like to understand the connection between our phase diagram and the tricritical quantum critical point 
phase diagram studied by Li et al~\cite{prl:stringari}.
Finally, as we have noted, our analysis of the Bogoliubov sound velocity neglected the effect of a harmonic trapping potential that is often present; although 
we expect
this to be qualitatively valid, an essential extension will be to properly account for the trapping potential. 

We gratefully acknowledge useful discussions with I. Spielman and A. Fetter.  
This work was supported by the Louisiana Board of Regents Grant LEQSF (2008-11)-RD-A-10 
and by the National Science Foundation Grant No. DMR-1151717.
This work was supported in part by the National Science Foundation under 
Grant No. PHYS-1066293 and the hospitality of the Aspen Center for Physics.

\appendix
\section{Effective low-energy Hamiltonian}
\label{Eq:appendix}

In this section we derive the low energy effective Hamiltonian for
a $^{87}$Rb BEC with spin-orbit coupling, focusing on states near
the minima of $\varepsilon_{-}(\bp)$ (occuring at $\pm p_0\approx k_L$  for $\delta\to 0$).
  Our analysis closely follows
Ref.~\onlinecite{nat:lin}.  We start by noting the eigenstates of the 
rotated Hamiltonian $\Hh_r$, Eq.~(\ref{Eq:rotated}): 
\bea
\psih_{\bp+} &=& 
\frac{1}{\curN(p_x,\delta)}
\begin{pmatrix}
-1
\\
f(p_x,\delta)
\end{pmatrix},
\\
\psih_{\bp-} &=& 
\frac{1}{\curN(p_x,\delta)}
\begin{pmatrix}
f(p_x,\delta)
\\
1
\end{pmatrix},
\eea
corresponding to the eigenvalues Eq.~(\ref{Eq:epspm}).  Here, 
 we defined 
\bea
&&\hspace{-1cm}f(p_x,\delta) =
\frac{ \delta+
\frac{2k_L p_x}{m} - \sqrt{ \Omega^2+\delta^2 +\frac{4k_L^2 p_x^2}{m^2}+\frac{4 k_L \delta p_x}{m} }}{\Omega},~~
\eea
and the normalization factor  $\curN(p_x,\delta) = \sqrt{1+f(p_x,\delta)^2}$.

 Next, we  
express the original field $\Psi(\br)$ in terms of operators $\psi_\alpha(\bp)$ with momentum $\bp$ in 
  band $\alpha = \pm$: 
\be
\label{Eq:psiexpansion2}
\Psi(\br) = \sum_{\bp,\alpha=\pm} \Psih_{\bp\alpha}(\br) \psi_{\alpha}(\bp),
\ee
where $\Psih_{\bp\alpha}(\br) = \Uh   \psih_{\bp\alpha}  {\rm e}^{i\bp \cdot \br}$ is the eigenfunction of $\Hh$.  At low
energies, it is sufficient to restrict attention to the lower ($-$) band and focus on  $\bp$ close to the 
 right $(\bp_r)$ and left $(\bp_\ell)$ minima of $\varepsilon_-(\bp,\delta)$:
\bea
\label{Eq:psiexpansion3}
\Psi(\br)\!&= &\! \sum_{p<\Lambda,a=r,\ell} \Psih_{\bp+\bp_a -}(\br) \psi_-(\bp+\bp_a) ,
\\
 &= & \sum_{p<\Lambda} \Big[\Psih_{\bp+\bp_r -}(\br)\psi_\downarrowp(\bp) +  
\Psi_{\bp+\bp_\ell -}(\br) \psi_\uparrowp(\bp)\Big],
\nonumber
\eea
where $\Lambda$ is a cutoff parameter, representing the range of momenta near the minima at $\bp_r$ and $\bp_\ell$
that are included in the sum.
In the second line of Eq.~(\ref{Eq:psiexpansion3}) 
we introduced the notation $\psi_\downarrowp(\bp) = \psi_-(\bp+\bp_r)$ 
and $\psi_\uparrowp(\bp)  =  \psi_-(\bp+\bp_\ell)$  for the states near $\bp_r$ and $\bp_\ell$; the notation
$\downarrowp$ and $\uparrowp$  follows since, for vanishing light-atom coupling $\Omega \to 0$,
the states near the right (left) minimum map onto the $\downarrow$ ($\uparrow$) 
band of Eq.~(\ref{Eq:hhat}).

 Plugging this into the single-particle Hamiltonian $\curH_0$, and using the orthonormality
of the eigenfunctions of Eq.~(\ref{Eq:hhat}), we obtain 
\be
 \curH_0 = \sum_{\bp<\Lambda,\sigma= \uparrowp,\downarrowp} 
\varepsilon_\sigma(\bp) \psi_\sigma^\dagger(\bp)\psi_\sigma^\phdag(\bp),
\label{Eq:simplifyapp}
\ee
where the dispersion $\varepsilon_\sigma(\bp)$ is given by $\varepsilon_\uparrowp(\bp) = \varepsilon_-(\bp+ \bp_\ell)$
and $\varepsilon_\downarrowp(\bp) = \varepsilon_-(\bp+ \bp_r)$.  

Equation~(\ref{Eq:simplifyapp}) can be simplified further by noting that, 
as shown below and in agreement with the expermental
findings of Ref.~\onlinecite{nat:lin},   the mixed phase is only 
stable for a small range of $\delta$ values, so that this parameter can be taken to be 
small.  To leading order in small $\delta/4E_L$, the minima of $\varepsilon_-(\bp)$
occur at
\be
p_{r,\ell} \simeq \pm k_L\sqrt{1-\Omegah^2} + k_L\deltah\frac{\Omegah^2}{1-\Omegah^2},
\ee
with the $+$ ($-$) corresponding to the right (left) minimum.  Here,
 we defined $\Omegah = \Omega/4E_L$ and $\deltah= \delta/4E_L$.
 Since stability of the mixed phase also requires $\Omegah \ll 1$ as
well as  $\deltah\ll 1$, it is clear that the final term in this expression can be neglected compared to
the first term, implying that the locations of the minima of $\varepsilon_-(\bp)$ are close
to $p_x = \pm k_L\sqrt{1-\Omegah^2}$.  Inserting these values into $\varepsilon_-(\bp)$, and again neglecting
terms of order $\Omegah^2 \deltah$, we find the energies of the local minima to be:
\be
\label{Eq:dropconstant}
\varepsilon_-(\bp_{r/\ell}) = E_L \big( -\Omegah^2 \pm 2\deltah), 
\ee
with the $-$ ($+$) corresponding to the right (left) minima.   The preceding calculations show that, for sufficiently small 
values of $\delta$, the effect of nonzero $\delta$ is simply to apply a chemical
potential difference, lowering the $\downarrowp$ state energy for $\delta>0$ and the $\uparrowp$ state energy for $\delta<0$. 
Expanding the dispersions $\varepsilon_{-}(\bp)$  to leading order  $\bp$ near these minima, we finally arrive at 
(including a chemical potential $\mu$ that couples to the density and defining $\mu_\uparrowp = \mu -\frac{1}{2}\delta$
and $\mu_\downarrowp = \mu +\frac{1}{2}\delta$):
\be
\label{Eq:simplify2app}
\curH_0 = \sum_{\sigma =\uparrowp,\downarrowp}\int d^3 r \big( \varepsilon(\bp) - \mu_\sigma) \psi^\dagger_\sigma(\br) \psi^\phdag_\sigma(\br).
\ee
In Eq.~(\ref{Eq:simplify2app}) we dropped an overall constant from the first term in Eq.~(\ref{Eq:dropconstant}).
Here,  the effective dispersion is 
\be
\label{Eq:effectivedispersionapp}
\varepsilon(\bp) =\frac{1}{2m^*} p_x^2 + \frac{1}{2m}(p_y^2+p_z^2),
\ee
 with a different
effective mass $m^*$ in the $x$ direction, reflecting the curvature of the minima of $\varepsilon_-(\bp)$, 
that satisfies $(m^*)^{-1} = m^{-1} (1-\Omegah^2)$.

The final single particle Hamiltonian Eq.~(\ref{Eq:simplify2}) possesses an exact degeneracy, at $\delta = 0$,
among the $\uparrowp$ and $\downarrowp$ states; however the interaction Hamiltonian does not possess this symmetry.
Indeed, as discussed in the main text, this is because of the spin-dependence of the $^{87}$Rb interactions,
captured by the Hamiltonian:
\bea
\hspace{-.25cm}\curH_1 = \frac{1}{2}\int d^3r \!\Big[(c_0 \!+\! c_2)\rho_\downarrow^2 \!+\!c_0 \rho_\uparrow^2\!+\!
2(c_0 \!+\! c_2)\rho_\uparrow\rho_\downarrow\Big],~~
\label{eq:honeapp}
\eea
where $\rho_\sigma=\Psi^\dagger_\sigma\Psi_\sigma$ with $\sigma=\uparrow, \downarrow$ and normal ordering is implied.  Since 
$c_0>0$  and $c_2<0$ with $|c_2|\ll c_0$, the  $\uparrow$ bosons having a larger intraspecies repulsion than the $\downarrow$ bosons.

To obtain the effective interactions among the dressed bosons, 
 we need to use Eq.~(\ref{Eq:psiexpansion3}) in Eq.~(\ref{eq:honeapp}).
For Eq.~(\ref{Eq:psiexpansion3}), we need the eigenfunctions near the minima at $\bp_r$ and $\bp_\ell$.  Approximating
the function $f(\bp +\bp_r)\simeq f(\bp_r)$  (and similarly for $f(\bp+\bp_\ell)$) in this formula, 
and defining the Fourier transform $\psi_\sigma(\br) = \sum_\bp {\rm e}^{i\bp\cdot \br} \psi_\sigma(\bp)$ (essentially taking
the cutoff parameter $\Lambda \to \infty$), we obtain
\bea
\label{Eq:psiup}
&&\Psi(\br) \simeq \frac{1}{\curN(\bp_r)}
\begin{pmatrix}
f(\bp_r){\rm e}^{2ik_Lx } 
\\
1
\end{pmatrix}  \psi_\downarrowp(\br)
\\
&& \qquad  \qquad  \qquad  \qquad  \qquad  \qquad 
+ 
\frac{1}{\curN(\bp_\ell)}
\begin{pmatrix}
f(\bp_\ell)
\\
{\rm e}^{-2ik_Lx} 
\end{pmatrix}
\psi_\uparrowp(\br).
\nonumber
\eea
Again focusing on the limit of small $\Omegah$,  we keep terms up to order $c_0\Omegah^2$ (discarding
terms with rapidly-varying exponential factors) and 
take the limit $\Omegah \to 0$ in the terms proportional to $c_2$ (since $|c_2|\ll c_0$).  As we found
for $\curH_0$, the corrections due to $\deltah$ are also subdominant, leading to the final
interaction Hamiltonian 
\bea
&&\curH_1 = \frac{1}{2}\int d^3r \Big[(c_0 + c_2)|\psi_\downarrowp(\br)|^4 +c_0 |\psi_\uparrowp(\br)|^4 
\nonumber \\
&&\quad +
2\big[ c_0(1+\Omegah^2)+c_2  \big]|\psi_\uparrowp(\br)|^2|\psi_\downarrowp(\br)|^2\Big],
\label{eq:lowenergyhoneapp}
\eea
where normal ordering is implied.  Thus, we see that, in agreement with Ref.~\onlinecite{nat:lin}, the leading
impact of SOC on the $^{87}$Rb interactions is to renormalize the interatomic interactions.

\end{document}